\documentclass{article}
\usepackage{graphicx} 
\usepackage[backend=biber, style=nature]{biblatex}
\usepackage{siunitx}
\DeclareSIUnit\torr{Torr}
\DeclareSIUnit\angstrom{\text{Å}}
\usepackage{amsmath}
\usepackage[margin=1in]{geometry}
\usepackage{tabularx}
\usepackage{booktabs} 
\usepackage{miller}

\title{Grain-size distribution characterization from single-shot x-ray diffraction using hierarchical Bayesian multi-emitter analysis}
\author{Hyun Sang Park, Gannon Murray, Rohit Berlia, \\ David Elbert, Timothy P. Weihs, Todd C. Hufnagel}
\date{September 2026}

\begin{document}
\maketitle

\begin{abstract}
The distribution of grain sizes in a polycrystalline material can be determined from a 2D x-ray diffraction pattern by counting and characterizing diffraction spots, so long as the spots are discrete. This approach breaks down if there is substantial overlap of the reflections due to either a large number of grains in the scattering volume or crowding of spots caused by crystallographic texture. As a result, the spot-counting approach to grain size determination is infeasible for many situations of practical interest.

We address this limitation by applying hierarchical probabilistic learning using a reversible-jump Markov chain Monte Carlo framework to jointly infer diffraction spot populations and their intensity distributions. Using synthetic diffraction patterns, we show how this hierarchical model allows recovery of diffraction peak intensities across a wide variety of grain-size distributions and crystallographic textures. From the recovered peak intensities we infer grain-size distributions themselves and characterize the regimes of grain size and texture for which this approach is practical. Finally, we demonstrate the application of our technique by comparing grain-size distributions in Cu-Ti alloy specimens obtained from 2D x-ray diffraction patterns with those from electron backscatter diffraction (EBSD). This method of grain-size distribution determination is expected to be particularly useful in the context of high-throughput experimentation, where the minimal sample preparation requirements and short exposure times of single-shot x-ray diffraction are substantial advantages relative to microscopy techniques.
\end{abstract}

\section{Introduction}
The properties of materials are heavily influenced by microstructure. This is especially true for mechanical properties of structural materials, where differences in grain-size distributions (GSDs) can strongly influence material response~\cite{miracle_emerging_2021}. The standard technique for measuring GSDs is electron backscatter diffraction (EBSD), which provides a detailed crystallographic map of a planar section through the sample. But EBSD also has some limitations, including the need for careful sample preparation and relatively long acquisition times for large areas~\cite{farrer_ebsd_2000}. 

X-ray diffraction can also be used to measure grain size. The well-known Scherrer equation allows determination of an average grain size from peak broadening, but only works for small grains (typically $<\qty{200}{nm}$)~\cite{holzwarth_scherrer_2011}. While grain size and shape can be mapped for larger grains with near-field high-energy diffraction microscopy (nf-HEDM) and diffraction contrast tomography (DCT), both require time-consuming collection of hundreds or thousands of diffraction patterns as well as sophisticated reconstruction algorithms, and nf-HEDM is limited to synchrotron sources. DCT can be performed in the laboratory but struggles to reliably measure the shape of grains smaller than about \qty{40}{\micro m}~\cite{fang_improved_2021, fang_implementation_2023,mcdonald_tracking_2021,bachmann_3d_2019}. There is thus a need for a technique to measure GSDs quickly, reliably, and over a wider range of grain sizes. This is especially relevant for automated, high-throughput laboratories, which are intended to rapidly gather large data sets for training artificial intelligence and machine learning models, for example for autonomous materials exploration and optimization campaigns~\cite{hufnagel_aimd-l_2026}.

An old idea for measuring grain size via x-ray diffraction relies on azimuthal variations in intensity around Debye-Scherrer cones~\cite{warren_xray_1960}. These variations are due to the number of crystals in the illuminated volume. If the number is small, the diffraction spots on a given Debye-Scherrer ring are discrete and easily counted. With decreasing grain size (and thus more crystals in the gauge volume) there is increasing spot overlap until finally the rings become continuous with little or no intensity variation. Calibrated and statistical relationships between the spottiness of these diffraction rings and microstructure have enabled researchers to characterize grain size, and in some cases grain-size distributions, from XRD through blackening measurements~\cite{hirsch_determination_1954}, signal thresholding~\cite{he_method_2011}, rocking curves~\cite{he_crystallite_2022}, variance statistics~\cite{yager_metrics_2014}, and other hybrid approaches~\cite{solayman_advanced_2025}. These methods have been successfully used for characterizing grain size in geologic materials~\cite{bramble_grain_2015} and pharmaceutical drugs~\cite{thakral_estimation_2018}.
A persistent challenge in all of these approaches is spot overlap, due either to a large number of crystals in the scattering volume or to crystallographic texture, which can make diffraction spots dense in certain regions of a Debye-Scherrer ring. Both of these make distinguishing the contributions of individual grains to the diffraction pattern difficult.

\begin{figure}
    \centering
    \includegraphics[width=1\linewidth]{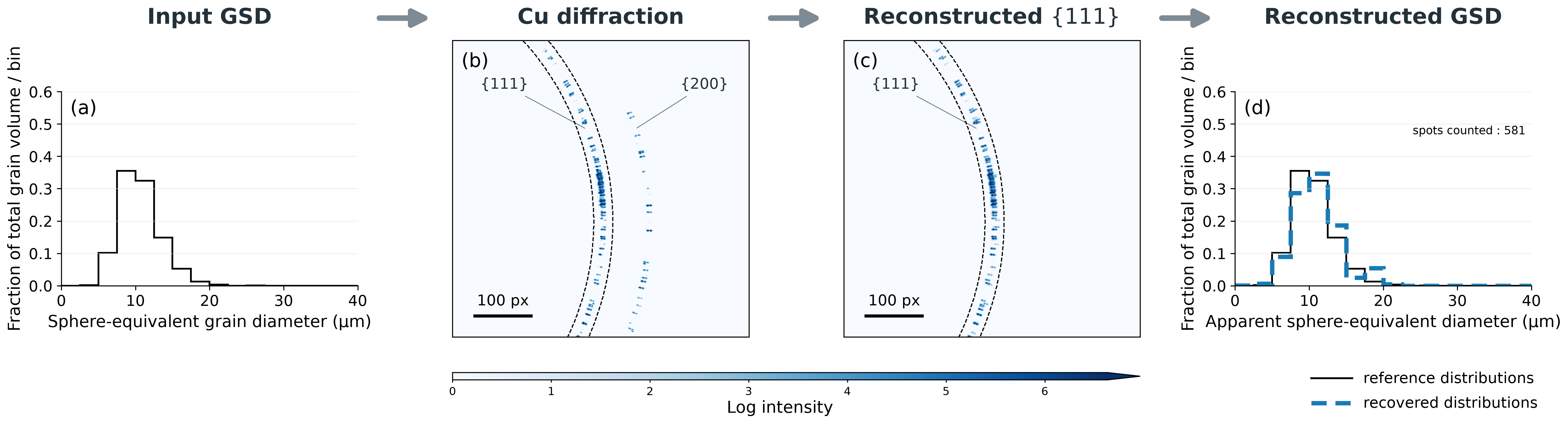}
    \caption{Workflow demonstrating grain-size distribution (GSD) recovery from synthetic 2D XRD data. We begin with a target GSD (a) which is sampled repeatedly to calculate the corresponding XRD pattern (b). Using the RJMCMC algorithm we reconstruct the intensities of the diffraction spots (c), producing a histogram of spot intensities (not shown) from which we infer the grain-size distribution (d). The process for recovering GSDs from experimental data is similar, except that we begin with the measured XRD pattern. (Diffraction patterns here show a portion of the \hkl{111} ring for simulated diffraction from polycrystalline copper.)}
    \label{fig:workflow}
\end{figure}

We address this challenge by introducing a hierarchical probabilistic learning framework based on reversible-jump Markov chain Monte Carlo (RJMCMC) sampling~\cite{green_reversible_1995}. Our approach draws on Bayesian multi-emitter analysis in single-molecule localization microscopy, where RJMCMC has been used to infer unknown emitter populations in densely populated intensity maps~\cite{fazel_bayesian_2019, van_dijk_bayesian_2023}. The basic approach is summarized in Fig.~\ref{fig:workflow} (with details in Section~\ref{sec:methods}) for trials involving synthetic diffraction patterns. We specify a microstructure (in terms of a GSD and orientation distribution function, ODF) and a number of grains in the scattering volume, and calculate the resulting diffraction pattern and corresponding intensity distribution. We then use RJMCMC to recover the spot-wise intensity distribution, and then convert that to a grain volume distribution.

To accommodate a broad range of potentially multi-modal spot-intensity distributions (resulting from different grain-size populations in a single specimen) with variable spot crowding, we use a shared, jointly learned intensity distribution prior whose parameters are jointly inferred with the number, positions, and intensities of diffraction spots. Joint inference enables intensity information to be shared across the diffraction ring, meaning well-resolved spots can help constrain intensity distribution to inform decomposition in areas with high spot crowding \cite{brewer_mapping_2009,feder_multiband_2020}. We represent this prior as a mixture of finite mixtures \cite{miller_mixture_2018} with up to three lognormal components. The number of components is also inferred with the jointly learned model, enabling the complexity of the intensity distribution to match the observed data. 

We assess the performance of this jointly learned intensity model on diffraction patterns generated from physically realistic grain-size distributions on simulated XRD patterns over a broad range of textures and grain-size distributions, establishing regimes where the RJMCMC approach can adequately recover accurate grain-size distributions. Following this, we evaluate the experimental application of our framework by comparing grain-size distributions recovered using the RJMCMC approach on experimentally measured XRD patterns from Cu-Ti alloy foils with results obtained using EBSD. The process is similar to that in Fig.~\ref{fig:workflow} except that we begin with the measured XRD patterns rather than synthetic patterns.

\section{Results}
\subsection{Intensity and grain-size distribution recovery in simulated data}
\begin{figure}
    \centering
    \includegraphics[width=1\linewidth]{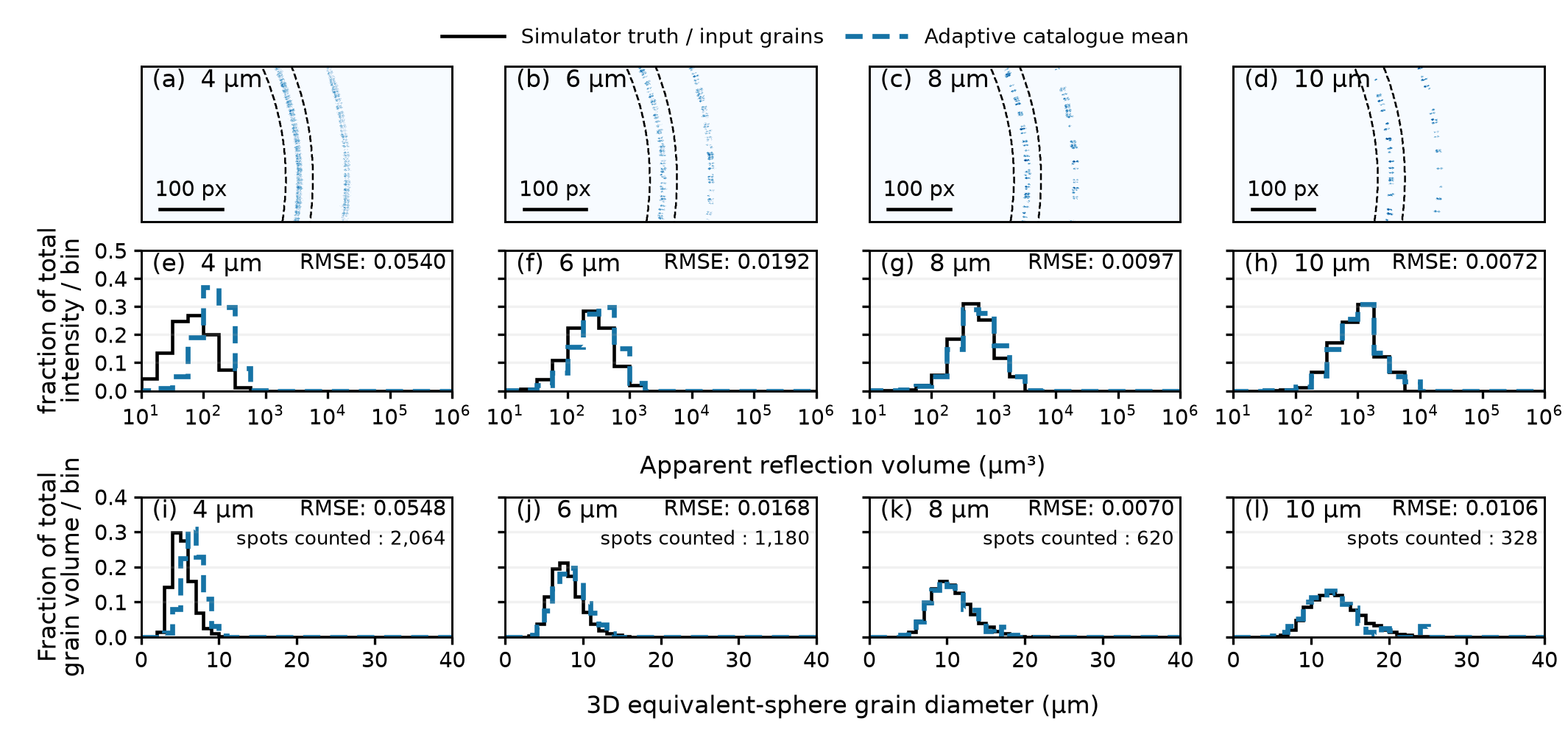}
    \caption{Selected cases demonstrating tested synthetic lognormal grain-size distributions with modes between \qty{4}{\micro\meter} and \qty{10}{\micro\meter} generated by the forward simulator. Experimental conditions here are optimized for smaller grain recovery (\qty{250}{\micro\meter} diameter X-ray spot). All patterns are generated with random crystallographic orientations (1 MRD).}
    \label{fig:size_small}
\end{figure}

The first step in the GSD analysis of a diffraction pattern is recovery of the diffraction spot intensity distribution. Figure~\ref{fig:size_small} shows results for XRD patterns from synthetic microstructures with lognormal grain-size modes ranging from \qty{4}{\micro\meter} to \qty{10}{\micro\meter} and random orientations. Agreement between the ground-truth intensity distribution and recovered distribution improves with increasing average grain size. For the smallest grain sizes, diffraction spot overlap is severe and the RJMCMC method cannot accurately recover the intensity distribution. But in larger regimes (\qtyrange{6}{10}{\micro m}) the intensity and grain-size distribution are accurately recovered even though there is substantial overlap of the diffraction spots. For example, at a lognormal mode of \qty{6}{\micro\meter}, the mean nearest-neighbor spacing of the spots is approximately 0.786 times the nominal full widths at half-maximum (FWHM) and 74.7~\% of the spots have a neighbor within one FWHM, yet the root mean squared error (RMSE) of the intensity distribution recovery is less than 0.02 (fig. \ref{fig:size_small}~{b, f, j}). 

In these simulated diffraction patterns, we use the minimum X-ray spot size of our XRD instrument (\qty{250}{\micro\meter}) to minimize spot counts and reduce spot crowding. It is additionally possible to recover even smaller grains by characterizing smaller gauge volumes or using an $hkl$ diffraction ring with a lower multiplicity.

On the other end of the size scale, intensity recovery RMSE values are small for microstructures with larger average grain sizes, but this does not assure that grain-size distribution recovery is viable because the simulation produces very few spots on the detector. Recovery of GSDs from XRD patterns for materials with large grains may be limited by the small number of such grains in the scattering volume. This limitation may be overcome by sampling a larger volume of material or using multiple scans of the same sample at different locations (assuming the microstructure is homogeneous). 

\begin{figure}
    \centering    \includegraphics[width=1\linewidth]{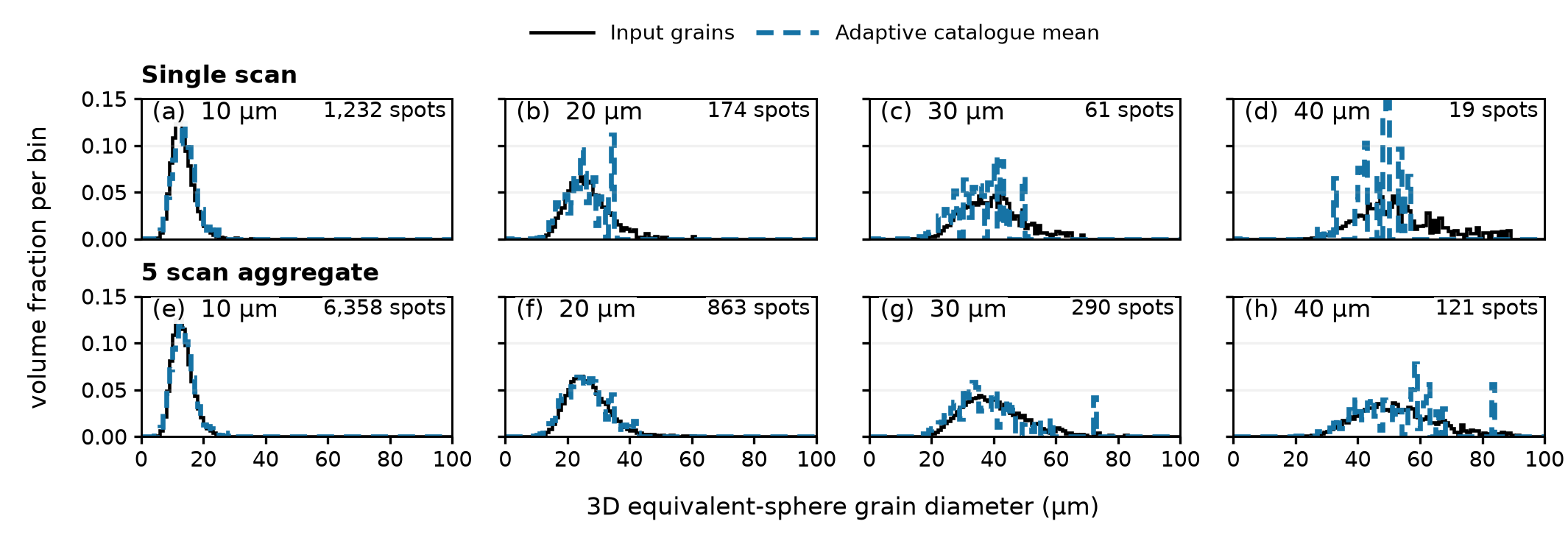}
    \caption{Selected cases demonstrating larger lognormal grain-size distributions with (a-d) a single diffraction pattern and (e-h) aggregates of five diffraction patterns. Experimental conditions are optimized for counting statistics and use a \qty{510}{\micro\meter} diameter X-ray spot.}
    \label{fig:large_grains}
\end{figure}

We demonstrate this by aggregating GSD from five independent synthetic XRD patterns generated with shared underlying grain statistics across a range of grain sizes~(fig.\ref{fig:large_grains}). To improve our grain counts, we additionally set our X-ray spot size to \qty{510}{\micro\meter}: the maximum spot size on the associated physical instrument~\cite{park_automated_2025}. Pooling results reduced the GSD RMSE when compared to single diffraction scans in all cases shown. In these larger grain regimes, histograms with greater than 1000 counted spots seem to show smooth recovery of the lognormal distribution. Compared to EBSD, accessing these experimental volume regimes through XRD by taking more scans is more efficient due to shorter scan times and the lack of surface preparation requirements. 

\begin{figure}
    \centering
     \includegraphics[width=1\linewidth]{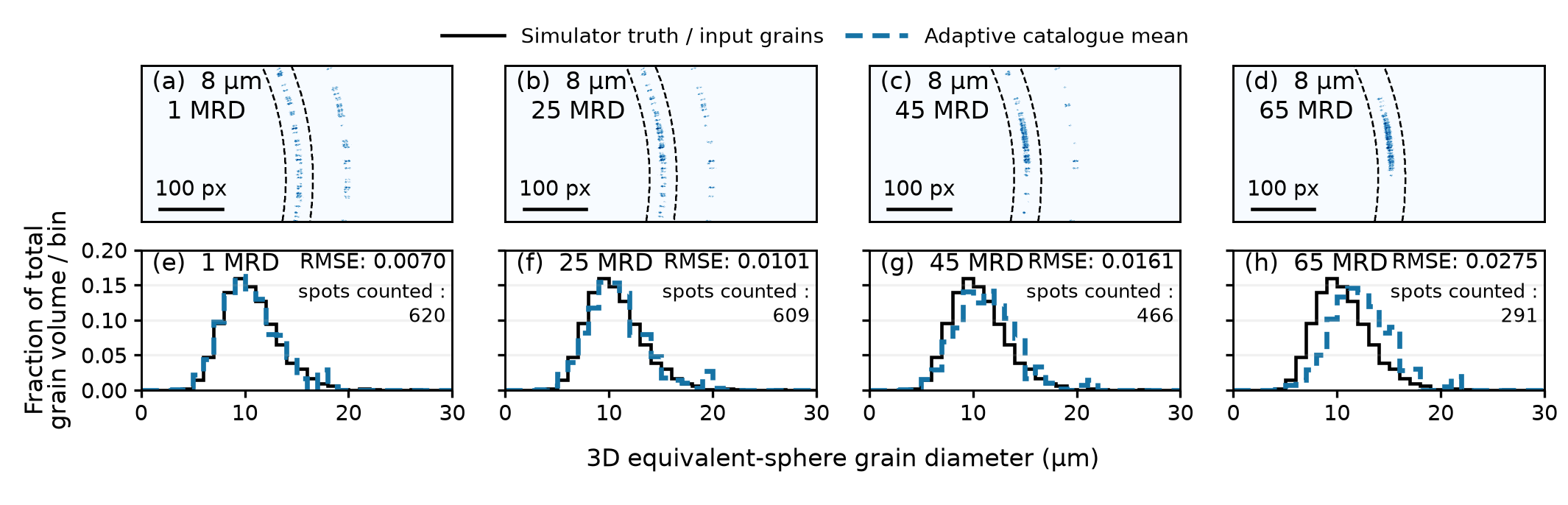}
    \caption{(e-h) Grain-size distribution recovery from (a-d) synthetic XRD patterns for a lognormal grain-size distribution mode of \qty{8}{\micro\meter}, with microstructures corresponding to four levels of \hkl[001] fiber texture represented in terms of multiples of random distribution (MRD). Grain-size recovery degrades for smaller grains and stronger texture due to diffraction spot overlap.}
    \label{fig:texture_shortened}
\end{figure}

Crystallographic texture can also challenge intensity recovery due to diffraction spot overlap, an effect similar to that of smaller grain sizes. In Figure~\ref{fig:texture_shortened} we show results for four grain populations that share a grain-size distribution across four levels of texture. We see that increasing the strength of the texture causes a shift in recovered GSD towards higher relative grain sizes in patterns with more spot overlap (\emph{i.e.} smaller grains or stronger texture). This behavior is consistent with unresolved intensity from weaker diffraction spots (smaller grains) being absorbed by brighter nearby peaks. 

\subsection{Experimental comparison with EBSD}
\begin{figure}
    \centering    \includegraphics[width=1\linewidth]{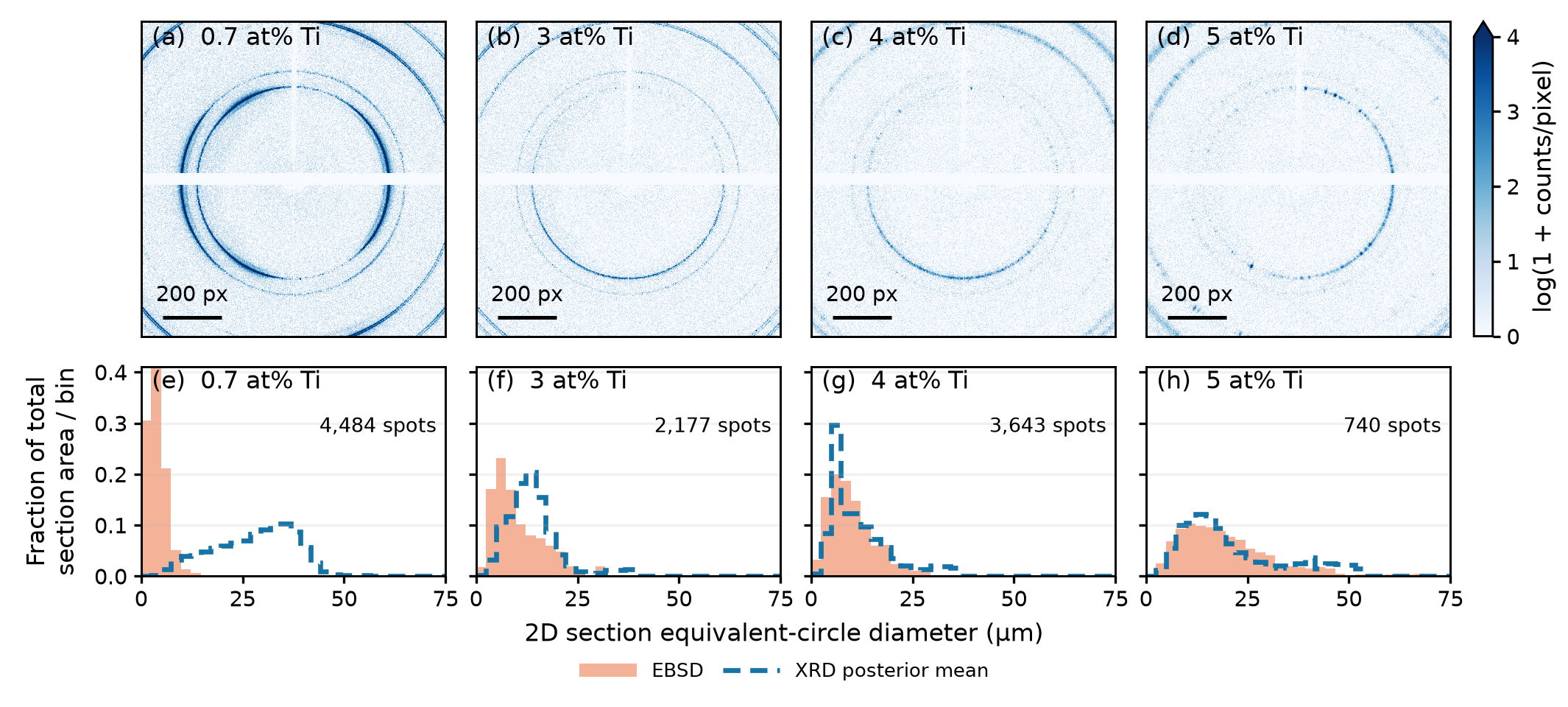}
    \caption{Analysis performed on experimentally measured X-ray diffraction patterns on sputtered, combinatorial Cu-Ti films \qty{200}{\micro\meter} thick. Panels (a-d) show Cu-Ti samples with 0.7, 3, 4, and 5 atomic percent Ti, respectively. GSD histograms obtained from EBSD data (red fill) are shown beneath their corresponding diffraction patterns (e-h). GSD histograms generated by the hierarchical Bayesian recovery method presented in this paper (blue line) are based on analysis of the \hkl{111} ring and compared on the same histograms. Both distributions use \qty{2.45}{\micro\meter} bins.}
    \label{fig:experimental_ebsd}
\end{figure}

We validate our analysis method using experimentally-measured XRD patterns obtained from sputter-deposited combinatorial samples of Cu-Ti. These measurements were performed as part of a larger high-throughput study, where acquisition parameters were not optimized for spottiness or grain-size distribution recoverability~\cite{park_automated_2025}. Rather, these scans were optimized for lattice parameter characterization and phase-identification, so a large spot-size (\qty{370}{\micro\meter} diameter) was used with the explicit aim of sampling as many grains as possible. Figure~\ref{fig:experimental_ebsd} compares the XRD-derived GSD obtained here with those from EBSD measurements on the samples. We note that EBSD provides a GSD based on the area of each grain in a 2D cross-section; here, we converted the GSD from the XRD measurements (which are based on a volume fraction) to equivalent area-weighted fractions through a spherical section transformation~\cite{van_der_jagt_estimation_2025}.

The 0.7~at.\%~Ti sample (Fig.~\ref{fig:experimental_ebsd}a,e)  has the smallest grains (as measured by EBSD) and shows the largest discrepancy between our XRD-derived GSD and EBSD results. The small grain sizes and strong texture of this specimen are consistent with conditions for which grain sizes were over-predicted in analyzing synthetic data (Figures~\ref{fig:size_small} and \ref{fig:texture_shortened}). Similar effects are also apparent for the 3~at.\%~Ti sample, though to a lesser degree.

For the 4~at.\%~Ti sample, the agreement between the XRD and EBSD GSDs is better. Although it is not apparent from the figure, the integrated intensity around the \hkl{111} ring is about 22\% lower for the 4~at.\%~Ti sample than for the 3~at.\%~Ti sample, suggesting that the total volume of crystals contributing to the \hkl{111} ring is also lower, likely due to texture. A smaller volume of scattering crystals, with similar grain-size distribution, implies less spot crowding, enabling recovery of smaller grains in the 4~at.\%~Ti sample. Even closer agreement is observed for the 5~at.\%~Ti sample, with its larger average grain size. 

\subsection{Optimizing for recovery}
\label{Optimizing for recovery}
\begin{figure}
    \centering
    \includegraphics[width=1\linewidth]{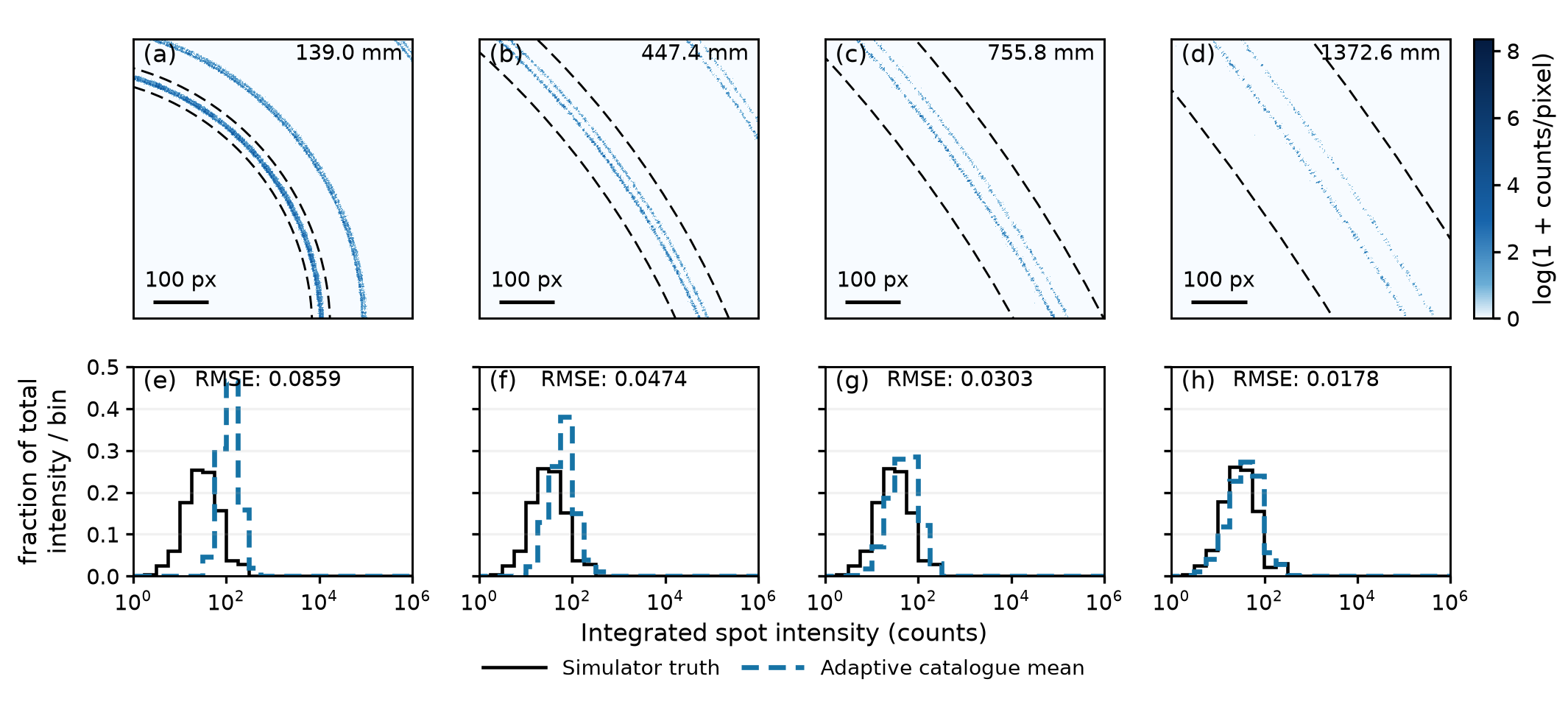}
    \caption{Simulated diffraction patterns, their intensity distributions (solid black line), and the recovered intensity distributions (dotted blue line) for a microstructure with a lognormal GSD with a \qty{3}{\micro\meter} mode and 1 MRD texture. Sample-to-detector distance is increased from (a,e) 139 mm to (b,f) 447.4 mm to (c,g) 755.8 mm to (d,h) 1372.6 mm while adjusting the diffraction pattern center to maintain the same region of interest while zooming in. Pixel widths of diffraction spots remain constant, meaning spot-center separation increases as the detector is moved further away from the sample.}
    \label{fig:sample_detector_distance}
\end{figure}

To a reasonable approximation, a given diffraction spot subtends a fixed solid angle on the detector. Thus, spot crowding can be reduced by increasing the distance between sample and detector, which effectively increases the angular resolution of the detector. We demonstrate this optimization for synthetic data in Figure~\ref{fig:sample_detector_distance}, which shows the same small segment of a diffraction ring for four detector distances. As expected, under a fixed pixel-width assumption, the histogram RMSE decreases monotonically with increasing detector distance. In a real experiment, of course, a tradeoff must be made between the increased spot separation on the detector on the one hand, and the inability to capture larger sections of the diffraction pattern.

There are other opportunities for optimizing recovery. For example, the number of observed spots is proportional to the multiplicity of the \hkl{hkl} planes for a given ring, so one can choose a ring for analysis that has a multiplicity better suited to the sample and experimental conditions. Furthermore, the gauge volume (and thus the number of diffracting crystals) can be adjusted by changing the beam size or (less conveniently) the sample thickness. 

\section{Discussion}
Our approach to grain-size distribution determination from single-shot XRD patterns is complementary to EBSD. On the one hand, EBSD provides a rich data set that includes information about grain morphology, crystallographic orientation, and relative orientations of adjacent grains, in addition to the grain-size distribution. On the other hand, EBSD requires careful sample preparation, is comparatively slow for mapping large areas, and is only sensitive to the near-surface region of the specimen. Single-shot x-ray diffraction, when performed in transmission with a high-energy x-ray beam, requires little specimen preparation, is fast ($\sim\qtyrange{1}{10}{s}$ per point with a laboratory system such as that in \cite{park_automated_2025}, or less at a synchrotron beamline), and samples the microstructure through the entire thickness of the specimen. It is therefore especially well-suited for high-throughput characterization of bulk materials.

Recovering diffraction spot intensities is crucial in estimating grain-size distributions from X-ray diffraction. Our hierarchical Bayesian approach, which jointly infers the intensity prior, extends previous approaches to intensity distribution recovery~\cite{he_method_2011, yager_metrics_2014,ingham_statistical_2014}. Joint inference removes the need to prescribe fixed intensity or grain-size expectation before running analysis. This is particularly useful for automated analysis in high-throughput settings, where grain populations and textures commonly vary between samples. Sweeps of synthetic XRD data across a wide range of grain-size regimes and texture strengths demonstrate the conditions under which our approach can adequately recover GSD, as well as those for which spot crowding biases the recovered intensity distributions. 
 
This intensity distribution recovery from X-ray diffraction patterns enables GSD characterization that shows good agreement with EBSD in the more favorable experimental cases. Here, we find similar constraints on practical GSD recovery: overlap associated with smaller grain size or texturing can cause over-prediction in XRD-derived GSD. Experimental design must therefore prioritize spot separability for intensity recovery. At the same time, however, recovering a grain-size distribution requires representative sampling of the microstructure. Performing analysis on a diffraction pattern with too few spots may improve spot intensity recoverability at the cost of grain sampling, increasing uncertainty in the final reconstructed grain-size distribution. A well-designed experiment must balance spot separability with adequate grain sampling for each diffraction pattern to support GSD analysis.

The presented multi-emitter analysis enables spot deconvolution in XRD patterns with overlapping and unevenly distributed diffraction spots. An additional advantage of this framework is that the number, locations, and intensities of diffraction spots are jointly learned with the parameters of a shared intensity distribution, reducing the need for preliminary spot fitting or calibration scans to design a spot intensity prior. Combining the demonstrated ability to accommodate crowding and texture, the adaptability of a jointly learned intensity prior, and the limited sample preparation required for XRD, the presented framework is a promising approach for autonomous and high-throughput microstructural characterization. 

\section{Methods}
\label{sec:methods}

\subsection{Synthetic diffraction data}
\label{sec:synthetic_data_methods}

We generate synthetic diffraction patterns using our \texttt{2D\_xray\_generator} forward XRD simulator~\cite{noauthor_harrison11252d_xray_generator_nodate}. Baseline simulations use parameters matching the physical configuration of our XRD instrument~\cite{park_automated_2025,hufnagel_aimd-l_2026}. For a prescribed gauge volume, grains are sampled until their cumulative equivalent volumes reach the target volume. Orientations are assigned independently from ODFs containing \hkl[001] fiber components and a random population. Texture strength is characterized in multiples of random distribution (MRD), ranging from 1 for random orientations to 65 in the strongest textures examined. The exact parameters used to generate simulated patterns for this study are presented in Section~\ref{supp:Simulated_data_parameters} 

Grain orientations determine reflection positions, while grain volumes, structure factors, polarization and excitation weights determine relative integrated intensities. The simulation then produces expected count scaling from which Poisson observations are drawn based on an input scan-time. The diffraction and spot intensity calculations are described in supplement~\ref{sec:synthetic_observations}.

\subsection{Bayesian intensity recovery and joint population learning}
\label{sec:joint_learning_methods}

We reconstruct diffraction spots from native detector counts using a Bayesian model that combines a fixed detector response, an additive background, and Poisson counting statistics. Each reflection family is analyzed independently, with detector geometry and response parameters fixed. All selected pixels within a ring enter a joint fit, allowing its components to share an intensity distribution (section~\ref{sec:components}).

Overlapping spots can admit several plausible combinations of component number, position and intensity. Reversible-jump Markov chain Monte Carlo (RJMCMC) explores these alternatives by adding, removing, splitting or merging diffraction components, alongside updates to their positions and intensities~\cite{green_reversible_1995}. Proposal frequencies and acceptance rules are described in section~\ref{sec:supp_sampling}.

We designed the intensity prior to accommodate broad and possibly multi-modal spot-intensity distributions using a finite mixture of lognormal populations. The number of active populations and their means, widths and weights are inferred jointly with component number, positions and intensities, following a finite-mixture formulation with an unknown population count~\cite{miller_mixture_2018}. The jointly learned intensity prior is an extension of prior design in studies that have used preliminary experiments with well-resolved peaks to calibrate intensity priors for later multi-emitter analysis \cite{fazel_bayesian_2019}. In XRD, the intensity prior is directly influenced by grain populations and diffraction conditions, meaning an intensity prior learned from a different XRD pattern may not transfer to further experiments. Learning occurs within each image--reflection fit without a separate supervised training stage. Better-resolved signal helps constrain the shared intensity distribution, which in turn informs alternative decompositions of crowded regions. This approach does, however, assume that sparsely populated and densely populated regions are populated by spots with a common intensity distribution. 

The objective of our framework is to recover intensity distributions within a practical sampling budget. Sampling used a prespecified budget of 150,000 draws, with periodic diagnostic checks described in section~\ref{sec:supp_sampling}. We expect a comprehensive joint-posterior exploration to be computationally demanding because unknowns in number, positions, and intensities of diffraction spots generating a large parameter space. Instead, we define convergence using Monte Carlo precision of the recovered GSD. The criterion requires the Monte Carlo standard error root-mean-square of the histogram bars to stay below 0.003 for two consecutive diagnostic checks. 

\subsection{Intensity to grain-size conversion}

For analysis, we use spectral doublets represented as paired Gaussian responses to represent our diffraction spots as we used in our synthetic pattern analysis. This is because the XRD source emits signals from In $K\alpha_1$ and $K\alpha_2$ wavelengths. The inferred full integrated intensities are converted to apparent component volumes using a reflection-specific volume/count coefficient, and then to spherical-equivalent diameters. This conversion is separate from intensity inference and depends on the exposure, calibration and visibility assumptions. The exact conversion and the conditional interpretation of synthetic size recovery are described in section~\ref{sec:calibrated_sizes}.

\subsection{Sample synthesis and independent EBSD comparison}
\label{sec:experimental_data_methods}

Free-standing Cu/Ti nanolaminate foils (\qty{200}{\micro\meter} thick) were fabricated by magnetron sputtering onto polished brass substrates using a custom-built, large-area sputtering system equipped with two oxygen-free high-conductivity (OFHC) Cu targets (99.99\% purity, Process Materials Inc.) and one shielded Ti target (99.95\% purity, Process Materials Inc.). A triangular shield on the Ti target established a controlled compositional gradient along the vertical direction of each substrate to achieve compositional variation from $\sim$0.1~at.~\% to $\sim$6.5~at.~\% Ti across the substrate vertically. Deposition was carried out at room temperature under an Ar sputtering pressure of \qty{1.8}{\milli\torr} following a base pressure of \qty{5e-6}{\torr}. The substrate carousel was rotated at \qty{2.5}{\hertz} to ensure circumferential uniformity, and the sputter powers were set at \qty{800}{\watt} for Cu and \qty{770}{\watt} for Ti.
After deposition, the foils were mechanically delaminated from their substrates and fully solutionized in a vacuum furnace (MRF High Temperature Furnace) with a base pressure of \qty{1e-6}{\torr} at \qty{900}{\celsius} for \qty{10}{\hour} followed by oil quenching. All specimens were heat treated together in a single batch to eliminate run-to-run variability in thermal history.

The EBSD specimens were mechanically polished using SiC P4000 sandpaper (sourced from Allied HighTech) followed by cloth polishing using \qty{0.05}{\micro\meter} colloidal silica suspension to achieve a mirror finish. Final surface preparation was carried out by ion beam milling using a Leica EM TIC 3X operated at 5 kV and 3 mA for \qty{10}{\minute} to remove any residual surface deformation layers.

EBSD scans were performed using a Tescan MIRA 3 GM field-emission scanning electron microscope (SEM) equipped with an Oxford Instruments Symmetry EBSD detector. The microscope was operated at an accelerating voltage of \qty{20}{\kilo\volt}. Orientation maps were acquired with step sizes ranging from \qtyrange{0.01}{2}{\micro\meter}, depending on the grain size of each specimen. Area-weighted grain-size distributions were determined using Oxford AZtecCrystal software, with grain size expressed as the equivalent circle diameter. For this analysis, high-angle grain boundaries (HAGBs) were defined using a misorientation threshold of \qty{15}{\degree}.

XRD characterization was performed on sputter-deposited Cu-Ti samples generated from the same batch but were not mechanically polished. Here, the sample uses a spot size of \qty{370}{\micro\meter}. The same intensity conversions utilized for analyzing synthetic microstructures are used to generate grain-volume distributions. 

\section{Acknowledgments}
Research was sponsored by the Army Research Laboratory and was accomplished under Cooperative Agreement Number W911NF-22-2-0121. The views and conclusions contained in this document are those of the authors and should not be interpreted as representing the official policies, either expressed or implied, of the Army Research Laboratory or the U.S. Government. The U.S. Government is authorized to reproduce and distribute reprints for Government purposes notwithstanding any copyright notation herein.

\section{Supplemental information}
\label{sec:supplemental_information}

\subsection{Synthetic diffraction and imaging model}
\label{sec:synthetic_observations}

\paragraph{Grain populations and orientation distributions.}
Grain-resolved patterns were constructed using \texttt{2D\_xray\_generator} simulator~\cite{noauthor_harrison11252d_xray_generator_nodate}. Each grain has an equivalent-volume diameter, a crystal-to-sample rotation, and a reciprocal lattice. In fixed-gauge-volume comparisons, the population size is coupled to the diameter distribution through the total grain volume.

The fiber-texture construction places a specified crystal direction near a sample direction and draws a uniform rotation about that direction. Mixtures with a random orientation population vary the texture strength. The reported \hkl[001] fiber series uses multiples of random distribution (MRD) to characterize concentration about the preferred axis. A pole-density value within a finite angular cone and the fraction assigned to a fiber component are distinct quantities. The orientation distribution and its MRD evaluation are retained with the realization. Grain orientations are represented by normalized quaternions and applied directly to crystal-frame reciprocal vectors.

\paragraph{Excitation, projection and instrumental response.}
For an incident wavevector $\mathbf k_0$, with $\lVert\mathbf k_0\rVert=1/\lambda$, and a rotated reciprocal vector $\mathbf g$, the excitation error is the distance from the Ewald surface,
\begin{equation}
\Delta_g=\left|\lVert\mathbf k_0+\mathbf g\rVert-\lVert\mathbf k_0\rVert\right|.
\label{eq:gs_excitation_error}
\end{equation}

To predict intensity as a function of volume, the simulation uses eqs.~2-3 by Kroon-Batenburg \cite{kroon-batenburg_accounting_2015}. These equations are further explained in Section~\ref{intensity_volume_map}. Outgoing rays are projected onto the detector plane. Relative integrated intensity is proportional to equivalent grain volume, squared basis structure factor, polarization and excitation weight~\cite{kroon-batenburg_accounting_2015}. The exact equations are presented in Section~\ref{intensity_volume_map}. Finite grain shape is represented by a volume-preserving ellipsoid whose rotated second moments are projected onto the detector and combined with the instrumental Gaussian response. The general renderer uses a radial Voigt profile, comprising Lorentzian size broadening and Gaussian instrumental, strain and projected-shape contributions, with a Gaussian tangential profile.

Study-specific geometry is retained with each configuration. Relative simulator intensities require exposure scaling before Poisson observations are drawn. The grain model does not enforce a space-filling microstructure or partial illumination of individual grains.

\subsection{Simulated data parameters}
\label{supp:Simulated_data_parameters}
\paragraph{Instrument parameters} Simulation conditions were originally designed to match the configuration of the XRD instrument used in this study~\cite{park_automated_2025}. Simulations were performed using an FCC Cu matrix with a lattice parameter of \qty{0.3613}{\nano\meter} using an Indium $K\alpha_1$ wavelength of \qty{0.05121251}{\nano\meter} and $K\alpha_2$ wavelength of \qty{0.05165571}{\nano\meter}. The detector consists of $1000\times1000$ pixels with side lengths of \qty{75}{\micro\meter} positioned \qty{138.98}{\milli\meter} from the sample. Simulated samples had a thickness of \qty{200}{\micro\meter}. The incident photon flux used is $5.10848\times10^6$ photons per second, matching a calibrated response from the physical instrument. We use a \qty{150}{\second} scan time and analyze the \hkl{111} ring in all cases unless otherwise stated. 

\paragraph{Simulated data parameters} The lognormal diameter population sweep for recovering smaller grain populations in figure~\ref{fig:size_small} generated 4, 6, 8, and \qty{10}{\micro\meter} lognormal distributions with a texture strength of 1 MRD (random distribution of orientations) with a spot size of \qty{250}{\micro\meter}. The lognormal diameter population sweep for recovering larger grain populations in figure~\ref{fig:large_grains} generated 5 XRD patterns each of 10, 20, 30, and \qty{40}{\micro\meter} lognormal distributions with a texture strength of 1 MRD (random distribution of orientations) with a spot size of \qty{510}{\micro\meter}. The texture sweep used a lognormal GSD with a mode of \qty{8}{\micro\meter} with texture strengths of 1, 25, 45, and 65 MRD to generate fig.~\ref{fig:texture_shortened}. All lognormal distributions use a standard deviation $\sigma_{\ln D}$ of 0.25

\paragraph{Sample-to-detector study} The sample-to-detector study (fig. \ref{fig:sample_detector_distance} generated simulated data with a lognormal GSD of \qty{3}{\micro\meter} and texture strength of 1 MRD, while adjusting the sample-to-detector distance from 139.0 to 447.4 to 755.8 to \qty{1372.6}{\milli\meter}. These sample-to-detector distances correspond with detector centers that were translated to keep the analyzed region in the middle of the detector.

\subsection{Joint Bayesian model of diffraction components}
\label{sec:components}
\paragraph{Native counts and joint ring analysis.}
Diffraction-spot analysis infers an unknown number of components from native detector counts, adapting Bayesian multiple-emitter approaches developed for single-molecule localization microscopy\cite{fazel_bayesian_2019,van_dijk_bayesian_2023}. Each reflection family is analyzed independently within an annular region bounded by scattering angle, $2\theta$. Annular bounds are prescribed or selected by maximizing observed counts within a fixed window. Detector geometry and response parameters are fixed.

All selected pixels enter a single likelihood with a shared intensity distribution and additive background. Azimuthal sectors define spatial count priors and diagnostic summaries. Component responses are evaluated across sector boundaries on every observed pixel within their support. Candidate centers occupy unit detector cells obtained by two iterations of nearest-neighbor dilation of the observed-pixel mask, bounded by the detector image. Each cell is assigned to a sector by its center azimuth. This support admits components just outside the observed annulus whose response tails contribute to the data.

\paragraph{Shared intensity populations.}
The jointly learned model has an unknown number $S$ of intensity populations, distinct from the $K$ spatial components. We assign $P(S=s)=1/4$ for $s=0,1,2,3$, with $K=0$ for $S=0$ and $K>0$ otherwise. For $S>0$, the conditional log-intensity density is
\begin{equation}
f_a(a\mid S,\boldsymbol\phi)
=\sum_{h=1}^{S}w_h\,\mathcal N(a;m_h,\sigma_h^2),
\qquad \sum_{h=1}^{S}w_h=1,
\label{eq:gs_intensity_prior}
\end{equation}
where $\boldsymbol\phi$ contains the population means, widths and weights. Each $m_h$ is uniform over explicitly supplied bounds $(m_{\min},m_{\max})$ in natural-log counts. The log standard deviation is uniform between $\log(0.24)$ and $\log(4.5)$, and the active weights have a Dirichlet$(1,\ldots,1)$ prior. The mean bounds constrain population centers rather than individual component intensities. Learning occurs within each image--reflection fit: population parameters and population number are inferred jointly with the component catalog throughout retained sampling, without a separate supervised training stage. The shared distribution couples intensity inference across the ring, without imposing a grain-volume constraint or a prior on physical grain diameter.

\paragraph{Spatial count and background priors.}
For sector $q$ with supported area $\Omega_q$ in native pixel units, the unconditioned abundance model is
\[
\lambda_q\sim\operatorname{Gamma}\!\left(
\alpha_c,\ \mathrm{rate}=\frac{\alpha_c}{\rho\Omega_q}\right),
\qquad
K_q\mid\lambda_q\sim\operatorname{Poisson}(\lambda_q).
\]
The latent rates are integrated out; before conditioning, sector counts are independent with mean $\rho\Omega_q$ and variance $\rho\Omega_q+(\rho\Omega_q)^2/\alpha_c$. Active models condition their joint count distribution on $K=\sum_qK_q>0$. Its excluded empty-catalog probability is $p_0=\prod_q[1+\rho\Omega_q/\alpha_c]^{-\alpha_c}$. Conditional on sector counts, component centers are uniform within the corresponding support cells. The supplied whole-ring recipe uses $\rho=0.005$ components per supported pixel and $\alpha_c=1$.

For exposure ratio $e>0$ relative to a declared reference exposure, the background prior places probability $1/2$ at $b=0$ and $1/2$ on an exponential distribution with rate $50/e$, truncated to $0<b\leq3$ counts per pixel. For a catalog ordered by $x$, the joint posterior is
\begin{equation}
\pi(S,\boldsymbol\phi,K,\boldsymbol\theta,b\mid\mathbf d)
\propto
\mathcal L\,p_b(b)\,p(S)\,p(\boldsymbol\phi\mid S)\,
g_S(K,\mathbf r)\prod_{j=1}^{K}f_a(a_j\mid S,\boldsymbol\phi).
\label{eq:gs_posterior}
\end{equation}
Here $g_S$ combines the conditioned sector-count law with the ordered-position density. For an active catalog the latter contributes $\prod_q K_q!\,\Omega_q^{-K_q}$; the count law is divided by $1-p_0$. For $S=0$, the catalog is empty and the intensity product is one. 

\subsection{RJMCMC sampling, diagnostics and reproducibility}
\label{sec:supp_sampling}

\paragraph{Reversible-jump sampling.}
Posterior sampling uses reversible-jump Markov chain Monte Carlo\cite{green_reversible_1995}. The component kernel proposes birth, death, position, log-intensity, split, merge and two-component redistribution moves with relative frequencies $2{:}2{:}5{:}4{:}2{:}2{:}1$. Births propose a center uniformly over the complete support and an intensity from the current population density. Deaths select a component uniformly. A first birth in the adaptive model also proposes $S$ uniformly from 1--3, while the last death enters the empty model. Their acceptance ratios include the conditional-count normalizer and model-selection probabilities. Position and log-intensity moves use Gaussian perturbations with scale multipliers $0.1$, $1$ or $5$, selected uniformly. Position proposals compensate intensity by the ratio of old to new observed response mass when both masses are positive. Nearby-pair selection retains nonzero probability for distant pairs. Each proposal is accepted with probability
\begin{equation}
\alpha=\min\!\left\{1,
\frac{\pi(\mathbf z'\mid\mathbf d)\,
q_{m'}(\mathbf u'\mid\mathbf z')}
{\pi(\mathbf z\mid\mathbf d)\,
q_m(\mathbf u\mid\mathbf z)}
\left|\frac{\partial(\mathbf z',\mathbf u')}
{\partial(\mathbf z,\mathbf u)}\right|\right\},
\label{eq:gs_rj_acceptance}
\end{equation}
where $\mathbf z$ denotes the complete state and $q_m$ includes the move probability, component or pair selection, and auxiliary-variable density. Reverse probabilities are evaluated in the proposed state.

Splitting preserves total intensity and the intensity-weighted centroid. For $f=\{1+\exp(-v)\}^{-1}$ and displacement $\boldsymbol\delta$, the daughter positions are $\mathbf r-(1-f)\boldsymbol\delta$ and $\mathbf r+f\boldsymbol\delta$, with log intensities $a+\log f$ and $a+\log(1-f)$. The auxiliary logit is Gaussian with standard deviation $1.5$; displacement is Gaussian in $y$ and half-Gaussian in positive $x$, with a supplied scale. The split and inverse merge have unit absolute Jacobian in these coordinates.

One background proposal follows each block of component proposals. It selects with equal probability a switch between zero and the positive slab or a log-background random walk within the slab. 
The hierarchical model uses three auxiliary population slots with proper independent priors, including inactive slots, so unused parameters integrate out. Population weights are obtained by normalizing independent exponential variables over active slots. Inactive slots are refreshed from their priors, and labels are permuted. Population birth uses an equal mixture of a prior draw and a proposal near an active population, with its forward/reverse density correction. An independent joint-prior proposal, attempted with probability $0.005$ per recorded-state update, draws the complete catalog and intensity populations; its acceptance ratio reduces to the likelihood ratio. Over-capacity proposals are rejected.

\paragraph{Sampling budgets and diagnostic criteria.}
\label{diagnostics}
Independent chains start from an empty catalog and a nonempty catalog near the prior expected count, subject to capacity. Budgets, component proposals per recorded state, catalog capacity and seeds are supplied explicitly. The packaged recipe uses two chains, 253,600 requested warm-up component proposals, 344 component proposals per recorded state, batches of 5,000 states, at most 150,000 retained states per chain, and a capacity of 8,192 components. Warm-up completes whole update blocks. Position, log-intensity and population proposal scales adapt during warm-up toward an acceptance rate of $0.3$ and are then fixed; the background log-walk scale remains $0.5$.
Strict sampling success uses Monte Carlo standard errors (MCSE) of the GSD~\cite{jones_fixed-width_2006}. A ranked set of the largest bins in the GSD totaling to 98~\% of the recovered volume is checked for root-mean-square per-bin MCSE. GSD histogram convergence, and therefore completion of our analysis, is defined by that root-mean-square falling below 0.003. 

The default first check occurs after 5,000 states per chain. Following a failed check, we check diagnostics for convergence at a 1,000-state interval until it reaches a cap at 150,000 draws. Two consecutive passes are required to determine convergence. Proposals exceeding component capacity flag the fit as unsuccessful.

\paragraph{Posterior summaries and reproducibility.}
Mean component count, population-number probabilities, and mean detector predictions use all retained states. Uniform reservoir sampling retains whole-ring catalogs, with a default of 64 configurations per chain, preserving chain and iteration identities. Histogram precision uses regular whole-fit fraction traces, with count-bin edges and retained-state stride recorded independently of the catalog reservoir. Population parameters are recorded at every retained state; compact traces are saved by default and full traces optionally. Components are retained throughout the supported region without an intensity reporting threshold.

The implementation, \texttt{grainsize-xrd} version 0.2.0, uses Python 3.11 or later, NumPy, SciPy, Numba and ArviZ. Chains and reflection fits run concurrently, with separate deterministic random streams for chain proposals and reservoir retention. NPZ arrays and JSON metadata record configurations, budgets and retained counts, catalog and input/source identities, diagnostic checks, histogram edges and cadence, per-bin precision metrics, unassessed bins and software versions.

\subsection{Recovery metrics and histogram construction}
\label{sec:recovery_metrics}

\paragraph{Component matching and recovery probability.}
Recovery is evaluated against the known detector-component catalog rather than every grain supplied to the simulator. For each retained configuration in the controlled benchmark, a one-to-one assignment minimizes the total Euclidean distance between reference and fitted component positions. Position recovery requires every reference component to have an assigned component within $0.25\,\mathrm{FWHM}_t$, where $\mathrm{FWHM}_t=\sqrt{8\log2}\,\sigma_t$. Joint position--intensity recovery additionally requires
\begin{equation}
\left|\frac{I_{\mathrm{fit},j}}{I_{\mathrm{true},j}}-1\right|\leq0.25
\label{eq:gs_intensity_recovery}
\end{equation}
for every matched component. Configurations with fewer fitted than reference components fail. Additional fitted components are permitted, so matching success is reported separately from the component-count posterior and mean owned-component count.

\paragraph{Posterior histogram construction.}
Histograms are constructed within each posterior draw. Joint-fit intensity summaries use all supported components regardless of strict sampling success. Packaged calibrated-size summaries use owned components from strictly successful local cores or whole-ring joint fits. For diameter bin $B_b$ and weight $w_j^{(s)}$, the fraction in draw $s$ is
\begin{equation}
f_b^{(s)}=
\frac{\sum_{j\in\mathcal E_s}w_j^{(s)}\,
\mathbf{1}\!\left\{D_j^{(s)}\in B_b\right\}}
{\sum_{j\in\mathcal E_s}w_j^{(s)}},
\label{eq:gs_histogram_fraction}
\end{equation}
where $\mathcal E_s$ is the component set appropriate to the summary. For whole-fit intensity precision, the same expression uses count bins with $I_j^{(s)}$ in place of diameter $D_j^{(s)}$ and weights $w_j^{(s)}=I_j^{(s)}$. Number-, projected-area- and volume-weighted fractions use $w_j=1$, $A_j$ and $V_j$, respectively. Histograms used for stopping criterion use \qty{1}{\micro\meter} bins through \qty{200}{\micro\meter} and a larger-diameter overflow bin. Normalization includes the full represented distribution, including overflow. Empty draws have undefined fractions. In strictly gated size outputs, failed cores represent missing coverage. Reference and fitted catalogs use the same bin edges and weighting. Reflection families remain separate rather than being pooled as observations of unique physical grains.

\subsection{Apparent component-size distributions}
\label{sec:calibrated_sizes}

\paragraph{Intensity-to-volume mapping.}
\label{intensity_volume_map}
The inferred intensity is the full integrated component response, including both lobes when a spectral doublet is used. Size conversion is applied once to this total, rather than separately to the lobes or to the counts remaining after cropping. For a supplied reflection-specific volume/count coefficient as a function of volume $C_h(V_j)$, apparent component volume, spherical-equivalent diameter and projected area are, respectively:
\begin{equation}
V_j=C_h(V_j)I_j,\qquad
D_j=\left(\frac{6V_j}{\pi}\right)^{1/3},\qquad
A_j=\frac{\pi D_j^2}{4}.
\label{eq:gs_size_conversion}
\end{equation}
Here $C_h$ is expressed in cubic micrometers per integrated count. It is external to component fitting and conditional on the exposure, reflection and calibration convention. 

For actual intensity to volume conversions, we invert eqs. 2-3 developed by Kroon-Batenburg~\cite{kroon-batenburg_accounting_2015}:

\begin{equation}
    C_h(V_j)
    =
    \frac{
    2 s\sin^2\theta
    }{
    I_0 F^2 p
    \left(\dfrac{e^2}{mc^2}\right)^2
    n^2\lambda^3
    \displaystyle\int_{\eta_{\min}}^{\eta_{\max}}
    \frac{\sin^2\!\left[sB(\theta-\eta)\right]}
     {\left[B(\theta-\eta)\right]^2}
    P(\eta\mid\eta_0)\,d\eta}
\end{equation}

where s is the number of scattering layers, $I_0$ is the counts per unit area, $F^2$ is the squared structure factor, p is the polarization factor, ${e^2}/{mc^2}$ is the classical electron radius in Gaussian units, n is the number of unit cells per unit volume, $\lambda$ is the wavelength, B is an angular interference parameter, $\theta$ is the Bragg angle, and $P(\eta\mid\eta_0)$ is the angular probability density of intensity diffraction based on an offset angle $(\theta - \eta_0)$. We generate a table of $C_h(V_j)$ values using this equation, then use a lookup table to generate volume values based on $I_j$ when converting intensities back to volume. 

In cases where the X-ray source is expected to have both $K\alpha_1$ and $K\alpha_2$ wavelengths, we ascribe $2/3$ of the flux to a $K\alpha_1$ incident beam and $1/3$ of the flux to a $K\alpha_2$ beam, then combine detector responses. 

\paragraph{Comparison with planar sections characterized with EBSD.}
A three-dimensional equivalent-diameter distribution and a two-dimensional EBSD section distribution describe different sampling measures\cite{van_der_jagt_estimation_2025}. The retained spherical-section calculation provides a conditional forward comparison. For parent diameter $D$, the probability of intersecting a random plane is proportional to $D$. Conditional on intersection, the plane offset $z$ is uniform on $[-D/2,D/2]$, giving section diameter
\begin{equation}
d_{\mathrm{sec}}=\sqrt{D^2-4z^2}.
\label{eq:gs_section_diameter}
\end{equation}
Section-area fractions weight these intersections by $\pi d_{\mathrm{sec}}^2/4$. The reference implementation evaluates the expected fractions analytically and can generate finite collections of independent intersections to represent sampling variability.

\printbibliography{}
\end{document}